%% file: main.tex
\documentclass[sigconf]{acmart}
\pdfoutput=1

\AtBeginDocument{%
  \providecommand\BibTeX{{%
    \normalfont B\kern-0.5em{\scshape i\kern-0.25em b}\kern-0.8em\TeX}}}

\copyrightyear{2022}
\acmYear{2022} 
\setcopyright{acmlicensed}
\acmConference[ICSE-NIER'22]{New Ideas and Emerging Results }{May 21--29, 2022}{Pittsburgh, PA, USA}
\acmBooktitle{New Ideas and Emerging Results (ICSE-NIER'22), May 21--29, 2022, Pittsburgh, PA, USA}
\acmPrice{15.00}
\acmDOI{10.1145/3510455.3512772} \acmISBN{978-1-4503-9224-2/22/05}

\input{aliases.tex}
\input{utils.tex}

\begin{document}

\title{Towards Mining OSS Skills from GitHub Activity}

\author{Jenny T. Liang}
\affiliation{%
  \institution{University of Washington}
  \city{Seattle}
  \state{Washington}
  \country{USA}
}
\email{jliang9@cs.washington.edu}
\orcid{0000-0001-6722-9959}

\author{Thomas Zimmermann}
\affiliation{%
  \institution{Microsoft Research}
  \city{Redmond}
  \state{Washington}
  \country{USA}}
\email{tzimmer@microsoft.com}

\author{Denae Ford}
\affiliation{%
  \institution{Microsoft Research}
  \city{Redmond}
  \state{Washington}
  \country{USA}}
\email{denae@microsoft.com}

\renewcommand{\shortauthors}{Liang et al.}

\begin{abstract}

  Open source software (OSS) development relies on diverse skill sets. 
  However, to our knowledge, there are no tools which detect OSS-related skills.
  In this paper, we present a novel method to detect OSS skills and prototype it in a tool called \toolname. Our approach relies on identifying relevant \emph{signals}, which are measurable activities or cues associated with a skill. Our tool detects how contributors 1) teach others to be involved in OSS projects, 2) show commitment towards an OSS project, 3) have knowledge in specific programming languages, and 4) are familiar with OSS practices. We then evaluate the tool by administering a survey to 455 OSS contributors.
  We demonstrate that \toolname yields promising results: it detects the presence of these skills with precision scores between 77\% to 97\%. 
  We also find that over 54\% of participants would display their high-proficiency skills.
  Our approach can be used to transform existing OSS experiences, such as identifying collaborators, matching mentors to mentees, and assigning project roles. Given the positive results and potential impact of our approach, we outline future research opportunities in interpreting and sharing OSS skills.
\end{abstract}

\ccsdesc[500]{Software and its engineering~Open source model}

\keywords{open source software, skills detection, mining software repositories}

\maketitle

\section{Introduction}
Constructing open source software (OSS) depends on contributors with a diverse set of skills. Each skill is vital to OSS: problem-solving skills allow software developers to build new features to address issues; organizational skills help software maintainers manage the moving parts of an OSS project; and communication skills enable writers to generate clear, concise documentation and facilitate collaboration. Unlike in software engineering, where programmers primarily write code, many OSS contributors provide equally valuable non-code contributions \cite{allcontributors2022, trinkenreich2020hidden}. While OSS-related skills include a subset of software engineering skills, contributors also work in contexts unique to OSS (e.g., wrangling contributors, identifying funding, consistently collaborating in a distributed form).

Despite the importance of skills related to OSS development, to our knowledge, there are no tools that currently exist which detect such skills. 
Related work in software engineering has developed techniques to detect specific software engineering skills, for example, Java programming skills \cite{bergersen2014construction} and general programming experience \cite{siegmund2014measuring}.
Meanwhile, other work detects programming-related skills by pulling data from version control systems such as \gh \cite{papoutsoglou2019extracting, greene2016cvexplorer, montandon2019identifying, mockus2002icse, montandon2021mining, hauff2015matching}. Montandon et al. showed that this data could help identify experts in OSS communities \cite{montandon2019identifying} and predict technical roles of \gh users \cite{montandon2021mining}. Other work has demonstrated that \gh data can be used to extract skills for job recommendations \cite{greene2016cvexplorer, hauff2015matching}. %

These studies have significantly advanced the field, but focus largely on a single topic: mining technical programming skills. Thus, a gap still remains in mining OSS expertise, which encapsulates both soft and hard skills \cite{vadlamani2020studying} across many roles aside from software engineering. Rather than simply focusing on technical software development skills, our work focuses on mining \gh data to detect both soft skills and hard skills, which are vital to OSS development.

In this paper, we introduce a method to detect OSS skills and implement it in a tool called \ADDED{\toolname (\textbf{D}etect\textbf{I}ng \textbf{SK}ills in \textbf{O}SS)}, with promising results. Our approach relies on identifying accurate \emph{signals}, which are measurable activities or cues associated with a skill used to identify the presence of having that skill. For example, having proficiency in a programming language could be measured using a certain number of lines of code written in that language \cite{bergersen2014construction}. The notion of signals follows prior work, such as Marlow et al., who identified cues that \gh users utilized to judge a contributor's coding ability, project-relevant skills, and personality \cite{marlow2013activity}. 

We discuss how we identify relevant signals (Section \ref{sec:SkillsSignalsModel}) and outline \toolname's features (Section \ref{sec:ImplementingTool}). We then present an evaluation of \toolname and its results (Section \ref{sec:Evaluation}). Finally, we discuss the implications (Section \ref{sec:Discussion}) and next steps for this research (Section \ref{sec:FuturePlans}).

\section{Identifying Signals for Skills}
\label{sec:SkillsSignalsModel}

\input{tables/skills-signals-table}
To develop \toolname, OSS skills and signals need to be identified. The first author extracted relevant \textbf{OSS skills} by reading prior literature on software engineering expertise \cite{baltes2018towards, ahmed2012evaluating, li2020distinguishes, papoutsoglou2017mining} and social factors of OSS \cite{dias2021makes, gousios2015work, marlow2013activity, steinmacher2015social, tsay2014influence}. These papers were selected to include a broad set of roles and contribution types. The first author read each manuscript and identified key skills. %

Next, we performed additional literature review to identify \textbf{signals} for each skill. In this work, \emph{signals are expressed as true or false statements} (i.e., the signal is either present or not). Potential signals were elicited through the nine papers to identify OSS skills \cite{dias2021makes, tsay2014influence, gousios2015work, steinmacher2015social, baltes2018towards, ahmed2012evaluating, li2020distinguishes, marlow2013activity, papoutsoglou2017mining} and relevant literature on mining OSS activity \cite{bergersen2014construction, hata2021github, chouchen2021anti, chatterjee2021aid, zhou2012make, lee2017one, eluri2021predicting, bao2019large, raman2020stress}. We read through each manuscript, identified potential signals for as many OSS skills as possible, and recorded the source of each signal. We finalized the signals from literature by converting each one as a true or false statement by defining thresholds (e.g., an activity happens at least $N$ times, an activity occurs with a frequency at least at the $N$th percentile). We then generated signals which seemed viable to compute and were representative of the skill. 
We designed our own signals for \skill{Is familiar with OSS practices} because it was cited often in literature \cite{ahmed2012evaluating, gousios2015work, steinmacher2015social}, but prior work did not define clear signals for this skill.

During this process, authors consulted three OSS community experts who made significant contributions to OSS and regularly work with OSS stakeholders for their profession. They provided insight on important skills and signals. Based on the experts' feedback and the frequency of the skills cited in literature, we reduced the final model to four skills, which is shown in Table \ref{tab:SkillsSignalsTable}.

\section{Automatically Detecting Skills}
\label{sec:ImplementingTool}
After developing a model of OSS skills and their associated signals (see Section \ref{sec:SkillsSignalsModel}), the first author implemented \toolname in Python to detect a contributor's OSS skills from the model based on \gh user data. We use the \gh GraphQL API \cite{github2021graphql}, \gh REST API \cite{github2021rest}, and GHTorrent \cite{gousios2012ghtorrent} as sources of \gh data.

Our tool rates skills on a zero to five scale, where zero represents no proficiency and five represents a high level of proficiency. %
We use the function $rate(N, M)$ to determine skill level, where $N$ represents the number of signals present and $M$ represents the total number of signals a skill has. This function weights each signal equally.

\[ 
    rate(N, M) =
    \begin{cases} 
        0 & \tfrac{N}{M} = 0 \\
        1 & 0 < \tfrac{N}{M} \leq 0.2 \\
        2 & 0.2 < \tfrac{N}{M} \leq 0.4 \\
        3 & 0.4 < \tfrac{N}{M} \leq 0.6 \\
        4 & 0.6 < \tfrac{N}{M} \leq 0.8 \\
        5 & 0.8 < \tfrac{N}{M} \leq 1.0 \\
   \end{cases}
\]

\subsubsection*{Selecting programming languages} 
Programming languages that the tool detected were selected based on if they were present in both the 2020 State of the Octoverse \cite{github2020state} and the 2020 Stack Overflow Developer Survey \cite{stackoverflow2020developer}. This resulted in 9 programming languages: C, C\#, Java, JavaScript, PHP, Python, Ruby, Shell, and TypeScript.

\subsubsection*{\ADDEDZ{Excluding third-party libraries as} user contributions} 
Contributors often included code from third-party libraries in their commits, which has also been shown in prior work \cite{lopes2017dejavu}. \ADDED{Code from third-party libraries thus should be excluded from contributors' mined contributions.} To that end, we identify popular package managers for each programming language from a Wikipedia article on popular package managers \cite{wikipedia2021list}. We compile a list of installation folder names across all the package managers. Next, we analyze the top folder names for each programming language. We use a list of OSS for social good (OSS4SG) projects (i.e., OSS projects which address a societal issue and target a specific community) from Huang et al. \cite{huang2021leaving} We download the contents of each repository in this list and record its file tree. From this, we identify files written in the language based on file extensions and extract the path from the repository folder. We then retrieve the top 50 folder names associated with each programming language. For each language, we exclude the entire file tree under a folder name from the analysis when: 1) the folder name corresponds to an installation location of the programming language's package manager(s), or 2) the folder name is in the top 50 folder names for the programming language and the list of installation folder names across all package managers.

\subsubsection{Computing distributions} 
Some signals rely on the user activity being at a certain percentile and thus depends on an underlying distribution to compare to. To compute this, we generate a list of OSS contributors by recording the top 30 contributors \ADDEDZ{per project} from Huang et al.'s list of 437 OSS4SG projects~\cite{huang2021leaving}. We randomly select 500 \gh users from this list. Each distribution is computed based on these 500 users' relevant activity for the distribution.

\section{Evaluation}
\label{sec:Evaluation}
\subsection{Design}
We designed a two-part survey to validate \toolname (see Section \ref{sec:ImplementingTool}):
\begin{enumerate}
    \item a Qualtrics survey where participants submitted anonymized responses and  
    \item a Microsoft Forms survey where participants submitted personal identifying information (PII). \ADDEDZ{This survey was displayed after the completion of the Qualtrics survey.}
\end{enumerate} \ADDEDZ{The survey was implemented in two parts so PII was linked only to a small number of questions.} Participants could choose to only take the Qualtrics survey \ADDEDZ{and not provide any PII}.
Topics in the Qualtrics survey included the importance of our tool's OSS skills and the participants' willingness to display their skills and ratings on \gh. Topics in the Microsoft Forms survey included \gh usernames and self-assessments of the skills from \toolname. 
Some survey questions are shown in Figure~\ref{fig:survey} and the full survey instrument is available as supplemental material~\cite{supplemental-materials}. 
We sent the survey to a subset of contributors who authored commits, opened issues and pull requests, or commented on others’ issues and pull requests from Huang et al.'s list of 1,079 OSS and OSS4SG projects \cite{huang2021leaving}. \ADDED{Our survey was sent to a total of 9,095 OSS contributors with a response rate of 5\%.} The Qualtrics survey received 455 responses. The Microsoft Forms survey received 386 responses, resulting in 316 valid usernames with public code contributions from merged pull requests. 
After completing the survey, participants could join a sweepstakes to win one of four \$100 Amazon.com gift cards.

\ADDEDZ{The \gh usernames allowed us to compare the skill self-assessments with the skills scores computed by \toolname.}
To evaluate \toolname, we focus on its precision---that is, when the participant is confident they have a skill, does \toolname agree? We used two measurements of precision: 1) the precision of users who had self-evaluated skill levels greater than 0 (i.e., detecting the presence of a skill) and 2) the precision of users who had self-evaluated skill levels greater than 3 (i.e., detecting moderate to high skill proficiency).

For the analysis of the survey, we only look at close-ended questions and use standard statistical analysis techniques. We report percentages on how frequently participants agreed or strongly agreed with a statement and how frequently participants said a skill was important or very important. This follows Kitchenham's and Pfleeger's best practices to analyze survey data~\cite{survey-guidelines}.

\begin{figure}
    \begin{tcolorbox}[left=-14pt,right=2pt,top=2pt,bottom=2pt]
    {\sc \hspace{16pt}Part 1: Qualtrics Survey Questions (anonymous)}
    \begin{itemize}
        \item \questiontext{Q16}
        \item \questiontext{Q34}
        \item \questiontext{Q35}
    \end{itemize}
    \end{tcolorbox}
    \vspace{0.5\baselineskip}
    \begin{tcolorbox}[left=-14pt,right=2pt,top=2pt,bottom=2pt]
    {\sc \hspace{16pt}Part 2: Microsoft Forms Survey Questions}
    \begin{itemize}
        \item \ADDEDZ{\questiontext{MF5}}
        \item \questiontext{MF6}
        \item \questiontext{MF7}
        \vspace{0.5\baselineskip}
    \end{itemize}
    \vspace{-0.5\baselineskip}
    \end{tcolorbox}
    \caption{A subset of the survey questions. The complete survey instrument is in the supplemental materials~\cite{supplemental-materials}. }
    \label{fig:survey}
\end{figure}

\subsection{Preliminary Results}
\input{tables/skills-importance}
\input{tables/merged}

\subsubsection{Skill importance} All the detected skills were important to participants; the results are shown in Table \ref{tab:SkillsImportanceTable}. A majority of participants found the soft skills (\skill{Teaches others to be involved in the OSS project}, \skill{Shows commitment towards the OSS project}) to be important. Notably, participants found soft skills more important than hard skills. \skill{Is familiar with OSS practices} was also rated as important by a majority of participants, but was less important than the soft skills.

\subsubsection{Displaying skills} All detected skills would be displayed by a majority of participants (see Table \ref{tab:MergedTable}). Participants were most willing to share \skill{Is familiar with OSS practices}, \skill{JavaScript}, and \skill{Python}. \skill{PHP} was the least popular skill to share. Overall, there was variation in how willing participants were to share certain skills. In the free response, some participants were excited by sharing skills and suggested new languages to detect (e.g., Golang, Rust) or expressed encouragement. Others expressed concerns about the impact of \toolname, questioning whether it should be deployed.

\subsubsection{Tool accuracy} \toolname's performance is displayed in Table \ref{tab:MergedTable}. We find that \toolname identifies the presence of OSS skills with impressive performance, with precision scores ranging from 77\% (\skill{Ruby}) to 97\% (\skill{Is familiar with OSS practices}, \skill{Teaches others to be involved in the OSS project}). However, the overall performance of \toolname drops while detecting  moderately skilled to expert users. It is also notable that the rating distributions for \skill{Teaches others to be involved in the OSS project}, \skill{Shows commitment towards the OSS project}, and \skill{Is familiar with OSS practices} are skewed positively in the survey.

\section{Discussion}
\label{sec:Discussion}

Our evaluation shows encouraging results for \toolname. Participants are most excited to display programming language-related skills, which our tool detects with reasonable confidence. However, the high importance placed on soft skills by our participants should not be overlooked. This is especially relevant with emerging neural models that generate code with high quality, such as \gh Copilot \cite{github2021copilot}. Given its importance, future versions of \toolname could be enhanced to more accurately detect soft skills.

\subsubsection*{Potential applications.} \ADDED{Our results indicate a potential future for skill detection approaches within OSS development and software engineering.} While skills underlie these activities, they are not widely supported in tooling. Skills detection methods such as ours could be applied in practice to transform existing experiences. For example, \toolname could assist project maintainers with OSS team formation by supporting a search engine for potential collaborators based on skills or automatically recommending contributors for particular OSS roles. Furthermore, \toolname's skill ratings could be used to identify a contributor's potential areas of growth or recommend mentors with the expertise for the contributor's professional goals. Skills could also be displayed publicly to \gh profiles---users could self-select skills they were proficient in, while the platform could display a verified badge for detected skills.

\subsubsection*{Limitations.} One limitation of our evaluation is that self-rated skills is a biased measure, as participants may systematically overestimate or underestimate their skills. Additionally, limitations in describing skills in a survey may cause mismatched expectations of a skill. These may contribute to lower evaluation scores. For example, the signals for \skill{Is familiar with OSS practices} was designed to be simple to compute and beginner friendly, but participants rated themselves more harshly on the skill. Thus, when designing experiences with automatic skills detection, transparency in how the skill rating is computed is paramount and should be communicated clearly.

\section{Future Plans}
\label{sec:FuturePlans}
Our preliminary results indicate that there are some promising directions for \toolname. However, additional steps are required to improve upon our current approach, which we outline below.

\smallskip
\emph{Identifying additional signals from practitioners.} We hope to interview OSS contributors to understand how they evaluate their peers' expertise for the skills our tool detects. This could generate new signals to improve the accuracy of our tool. 258 participants have agreed to be interviewed for this work.

\smallskip
\emph{Defining weights of the signals.} In practice, each signal is not equally weighted in predicting a contributor's skill. Since our current approach does not support this, one next step could use a linear regression model to examine the relationship between signals and participants' self-evaluated skill level and then use the resulting model weights to weigh each signal in our tool. 

\smallskip
\emph{Evaluating the tool with other measures of skill.} We hope to run an additional evaluation using data sources less impacted by personal bias. For example, soft skills may be evaluated using peer evaluations, while hard skills may be evaluated using skill tests. Previous work has administered skill tests to determine Java expertise \cite{bergersen2014construction}.

\section{Conclusion}
We present \toolname, a tool to detect OSS-related skills which identifies \emph{signals} (i.e., measurable activities or cues associated with the skill) and then computes them from \gh data. \toolname detects the following skills: \skill{Teaches others to be involved in the OSS project}, \skill{Shows commitment towards the OSS project}, \skill{Has knowledge in specific programming languages}, and \skill{Is familiar with OSS practices}. We demonstrate this approach yields positive results, as \toolname detects the presence of OSS skills with precision scores between 77\% to 97\%. Additionally, a near majority of participants find the detected skills important to OSS. We expect to improve the tool and perform more rigorous evaluations in the future. Future work could design tools to augment existing OSS experiences or improve upon our current approach. Our supplemental materials are publicly available at \cite{supplemental-materials}.

\begin{acks}
We thank our survey participants for their insight and Christian Bird, Mala Kumar, Victor Grau Serrat, and Lucy Harris for their feedback. Jenny T. Liang conducted this work for an internship at Microsoft Research's Software Analysis and Intelligence Group.
\end{acks}

\bibliographystyle{ACM-Reference-Format}
\bibliography{acmart}

\end{document}

%% file: aliases.tex
\usepackage{pgfplots}
\pgfplotsset{width=10cm,compat=1.9}

\usepackage{stackengine}

\usepackage{xcolor}
\definecolor{darkgreen}{rgb}{0.05,0.5,0.05}

\newcommand{\skill}[1]{\emph{#1}}

\newcommand{\gh}{GitHub\xspace}

\definecolor{HIGHLIGHTCOLOR}{rgb}{0,0,0}
\newcommand{\ADDED}[1]{{\protect\color{HIGHLIGHTCOLOR}#1}}

\newcommand{\ADDEDZ}[1]{{\protect\color{HIGHLIGHTCOLOR}#1}}

\newcommand{\toolname}{{\small
\fontfamily{qag}\selectfont{\textsc{Disko}}}\xspace}

%% file: utils.tex
\usepackage{xcolor}
\usepackage{colortbl}
\usepackage{amsmath}
\usepackage{xspace}
\usepackage{fontawesome}
\usepackage{tcolorbox}
\usepackage{multirow}
\usepackage{multicol}

\usepackage{lookup}
\include{lookup_questions}
\newcommand{\questiontext}[1]{\lookupGet{#1}\xspace}

%% file: lookup_questions.tex
\lookupPut{Q32}{Which of the following information about your skills would you display on GitHub to just yourself, your collaborators, and the public?}

\lookupPut{C3}{Do you understand and consent to these terms?}
\lookupPut{Q4_1}{Demographics and Background What is your gender? - Selected Choice - Woman}
\lookupPut{Q4_2}{Demographics and Background What is your gender? - Selected Choice - Man}
\lookupPut{Q4_3}{Demographics and Background What is your gender? - Selected Choice - Non-binary / gender diverse}
\lookupPut{Q4_4}{Demographics and Background What is your gender? - Selected Choice - Prefer to self-describe}
\lookupPut{Q4_5}{Demographics and Background What is your gender? - Selected Choice - Prefer not to answer}
\lookupPut{Q4_4_TEXT}{Demographics and Background What is your gender? - Prefer to self-describe - Text}
\lookupPut{Q5_1}{What is your ethnicity/origin you identify with (select all that apply)? - Selected Choice - White}
\lookupPut{Q5_2}{What is your ethnicity/origin you identify with (select all that apply)? - Selected Choice - Asian}
\lookupPut{Q5_3}{What is your ethnicity/origin you identify with (select all that apply)? - Selected Choice - Black or African}
\lookupPut{Q5_4}{What is your ethnicity/origin you identify with (select all that apply)? - Selected Choice - Hispanic or Latino}
\lookupPut{Q5_5}{What is your ethnicity/origin you identify with (select all that apply)? - Selected Choice - Middle Eastern}
\lookupPut{Q5_6}{What is your ethnicity/origin you identify with (select all that apply)? - Selected Choice - American Indian or Alaska Native}
\lookupPut{Q5_7}{What is your ethnicity/origin you identify with (select all that apply)? - Selected Choice - Native American or Other Pacific Islander}
\lookupPut{Q5_8}{What is your ethnicity/origin you identify with (select all that apply)? - Selected Choice - Other (please specify)}
\lookupPut{Q5_9}{What is your ethnicity/origin you identify with (select all that apply)? - Selected Choice - Prefer not to report}
\lookupPut{Q5_8_TEXT}{What is your ethnicity/origin you identify with (select all that apply)? - Other (please specify) - Text}
\lookupPut{Q6}{How many OSS projects have you worked on (give your best estimation)?}
\lookupPut{Q7_1}{Currently, what are your roles in the OSS projects you are most active in? - Selected Choice - Documentation writer}
\lookupPut{Q7_2}{Currently, what are your roles in the OSS projects you are most active in? - Selected Choice - Bug submitter}
\lookupPut{Q7_3}{Currently, what are your roles in the OSS projects you are most active in? - Selected Choice - Graphic artist}
\lookupPut{Q7_4}{Currently, what are your roles in the OSS projects you are most active in? - Selected Choice - Translator}
\lookupPut{Q7_5}{Currently, what are your roles in the OSS projects you are most active in? - Selected Choice - Coder}
\lookupPut{Q7_6}{Currently, what are your roles in the OSS projects you are most active in? - Selected Choice - Maintainer}
\lookupPut{Q7_7}{Currently, what are your roles in the OSS projects you are most active in? - Selected Choice - Porter}
\lookupPut{Q7_8}{Currently, what are your roles in the OSS projects you are most active in? - Selected Choice - Tool builder}
\lookupPut{Q7_9}{Currently, what are your roles in the OSS projects you are most active in? - Selected Choice - GUI designer}
\lookupPut{Q7_10}{Currently, what are your roles in the OSS projects you are most active in? - Selected Choice - Other (please specify)}
\lookupPut{Q7_10_TEXT}{Currently, what are your roles in the OSS projects you are most active in? - Other (please specify) - Text}
\lookupPut{Q8}{OSS projects for social good are projects where the outcome distinctly targets a community of people to overcome a societal issue.Do you think all OSS projects are for social good?}
\lookupPut{Q9}{Have you worked on an OSS project that is for social good?}
\lookupPut{Q10}{How do you define skills in the context of OSS? What are some examples of such skills?}
\lookupPut{Q11}{What are your strategies to grow your skills in the context of OSS?}
\lookupPut{Q12}{Do you use GitHub to display skills and experiences that are relevant for OSS development?}
\lookupPut{Q13_1}{How do you display your skills and experiences that are relevant for OSS development on GitHub? - Selected Choice - GitHub profile description}
\lookupPut{Q13_2}{How do you display your skills and experiences that are relevant for OSS development on GitHub? - Selected Choice - Actively contributing to OSS projects}
\lookupPut{Q13_3}{How do you display your skills and experiences that are relevant for OSS development on GitHub? - Selected Choice - Having public repositories}
\lookupPut{Q13_4}{How do you display your skills and experiences that are relevant for OSS development on GitHub? - Selected Choice - Linking social media accounts to GitHub profile}
\lookupPut{Q13_5}{How do you display your skills and experiences that are relevant for OSS development on GitHub? - Selected Choice - Linking personal website to GitHub profile}
\lookupPut{Q13_6}{How do you display your skills and experiences that are relevant for OSS development on GitHub? - Selected Choice - Other (please specify)}
\lookupPut{Q13_6_TEXT}{How do you display your skills and experiences that are relevant for OSS development on GitHub? - Other (please specify) - Text}
\lookupPut{Q14}{Do you use LinkedIn to display skills and experiences that are relevant for OSS development?}
\lookupPut{Q15_1}{How do you display skills and experiences that are relevant for OSS development on LinkedIn? - Selected Choice - LinkedIn experiences}
\lookupPut{Q15_2}{How do you display skills and experiences that are relevant for OSS development on LinkedIn? - Selected Choice - Descriptions to LinkedIn experiences}
\lookupPut{Q15_3}{How do you display skills and experiences that are relevant for OSS development on LinkedIn? - Selected Choice - Skills}
\lookupPut{Q15_4}{How do you display skills and experiences that are relevant for OSS development on LinkedIn? - Selected Choice - Endorsements on skills}
\lookupPut{Q15_5}{How do you display skills and experiences that are relevant for OSS development on LinkedIn? - Selected Choice - Skill quizzes}
\lookupPut{Q15_6}{How do you display skills and experiences that are relevant for OSS development on LinkedIn? - Selected Choice - Recommendations from colleagues}
\lookupPut{Q15_7}{How do you display skills and experiences that are relevant for OSS development on LinkedIn? - Selected Choice - Honors and awards}
\lookupPut{Q15_8}{How do you display skills and experiences that are relevant for OSS development on LinkedIn? - Selected Choice - Courses}
\lookupPut{Q15_9}{How do you display skills and experiences that are relevant for OSS development on LinkedIn? - Selected Choice - Projects}
\lookupPut{Q15_10}{How do you display skills and experiences that are relevant for OSS development on LinkedIn? - Selected Choice - Other (please specify)}
\lookupPut{Q15_10_TEXT}{How do you display skills and experiences that are relevant for OSS development on LinkedIn? - Other (please specify) - Text}
\lookupPut{Q16}{Rate the following skills based on their importance for OSS projects.}
\lookupPut{Q16_1}{Rate the following skills based on their importance for OSS projects. - Teaches others to be involved in the OSS project}
\lookupPut{Q16_2}{Rate the following skills based on their importance for OSS projects. - Has commitment to the OSS project}
\lookupPut{Q16_3}{Rate the following skills based on their importance for OSS projects. - Has experience with specific tools, frameworks, or programming languages}
\lookupPut{Q16_4}{Rate the following skills based on their importance for OSS projects. - Has strong problem-solving skills}
\lookupPut{Q16_5}{Rate the following skills based on their importance for OSS projects. - Is familiar with OSS practices}
\lookupPut{Q16_6}{Rate the following skills based on their importance for OSS projects. - Puts in effort to make high-quality contributions}
\lookupPut{Q16_7}{Rate the following skills based on their importance for OSS projects. - Is a strong communicator}
\lookupPut{Q16_8}{Rate the following skills based on their importance for OSS projects. - Learns continuously}
\lookupPut{Q16_9}{Rate the following skills based on their importance for OSS projects. - Follows proper processes to make contributions to OSS projects}
\lookupPut{Q16_10}{Rate the following skills based on their importance for OSS projects. - Is a strong project manager}
\lookupPut{Q16_11}{Rate the following skills based on their importance for OSS projects. - Builds good relationships with stakeholders}
\lookupPut{Q16_12}{Rate the following skills based on their importance for OSS projects. - Has knowledge of algorithms, data structures, and programming paradigms}
\lookupPut{Q16_13}{Rate the following skills based on their importance for OSS projects. - Is available to answer questions}
\lookupPut{Q17_1}{Rate the following skills based on their importance for OSS projects that are for social good.OSS projects for social good are projects where the outcome distinctly targets a community of people to overcome a societal issue. - Teaches others to be involved in the OSS project}
\lookupPut{Q17_2}{Rate the following skills based on their importance for OSS projects that are for social good.OSS projects for social good are projects where the outcome distinctly targets a community of people to overcome a societal issue. - Has commitment to the OSS project}
\lookupPut{Q17_3}{Rate the following skills based on their importance for OSS projects that are for social good.OSS projects for social good are projects where the outcome distinctly targets a community of people to overcome a societal issue. - Has experience with specific tools, frameworks, or programming languages}
\lookupPut{Q17_4}{Rate the following skills based on their importance for OSS projects that are for social good.OSS projects for social good are projects where the outcome distinctly targets a community of people to overcome a societal issue. - Has strong problem-solving skills}
\lookupPut{Q17_5}{Rate the following skills based on their importance for OSS projects that are for social good.OSS projects for social good are projects where the outcome distinctly targets a community of people to overcome a societal issue. - Is familiar with OSS practices}
\lookupPut{Q17_6}{Rate the following skills based on their importance for OSS projects that are for social good.OSS projects for social good are projects where the outcome distinctly targets a community of people to overcome a societal issue. - Puts in effort to make high-quality contributions}
\lookupPut{Q17_7}{Rate the following skills based on their importance for OSS projects that are for social good.OSS projects for social good are projects where the outcome distinctly targets a community of people to overcome a societal issue. - Is a strong communicator}
\lookupPut{Q17_8}{Rate the following skills based on their importance for OSS projects that are for social good.OSS projects for social good are projects where the outcome distinctly targets a community of people to overcome a societal issue. - Learns continuously}
\lookupPut{Q17_9}{Rate the following skills based on their importance for OSS projects that are for social good.OSS projects for social good are projects where the outcome distinctly targets a community of people to overcome a societal issue. - Follows proper processes to make contributions to OSS projects}
\lookupPut{Q17_10}{Rate the following skills based on their importance for OSS projects that are for social good.OSS projects for social good are projects where the outcome distinctly targets a community of people to overcome a societal issue. - Is a strong project manager}
\lookupPut{Q17_11}{Rate the following skills based on their importance for OSS projects that are for social good.OSS projects for social good are projects where the outcome distinctly targets a community of people to overcome a societal issue. - Builds good relationships with stakeholders}
\lookupPut{Q17_12}{Rate the following skills based on their importance for OSS projects that are for social good.OSS projects for social good are projects where the outcome distinctly targets a community of people to overcome a societal issue. - Has knowledge of algorithms, data structures, and programming paradigms}
\lookupPut{Q17_13}{Rate the following skills based on their importance for OSS projects that are for social good.OSS projects for social good are projects where the outcome distinctly targets a community of people to overcome a societal issue. - Is available to answer questions}
\lookupPut{Q18}{List any skills not included in the questions above that you would display on your GitHub profile.}
\lookupPut{Q19}{Rate the following skills by how likely you are to improve them in the next 12 months.}
\lookupPut{Q19_1}{Rate the following skills by how likely you are to improve them in the next 12 months. - Teaches others to be involved in the OSS project}
\lookupPut{Q19_2}{Rate the following skills by how likely you are to improve them in the next 12 months. - Has commitment to the OSS project}
\lookupPut{Q19_3}{Rate the following skills by how likely you are to improve them in the next 12 months. - Has experience with specific tools, frameworks, or programming languages}
\lookupPut{Q19_4}{Rate the following skills by how likely you are to improve them in the next 12 months. - Has strong problem-solving skills}
\lookupPut{Q19_5}{Rate the following skills by how likely you are to improve them in the next 12 months. - Is familiar with OSS practices}
\lookupPut{Q19_6}{Rate the following skills by how likely you are to improve them in the next 12 months. - Puts in effort to make high-quality contributions}
\lookupPut{Q19_7}{Rate the following skills by how likely you are to improve them in the next 12 months. - Is a strong communicator}
\lookupPut{Q19_8}{Rate the following skills by how likely you are to improve them in the next 12 months. - Learns continuously}
\lookupPut{Q19_9}{Rate the following skills by how likely you are to improve them in the next 12 months. - Follows proper processes to make contributions to OSS projects}
\lookupPut{Q19_10}{Rate the following skills by how likely you are to improve them in the next 12 months. - Is a strong project manager}
\lookupPut{Q19_11}{Rate the following skills by how likely you are to improve them in the next 12 months. - Builds good relationships with stakeholders}
\lookupPut{Q19_12}{Rate the following skills by how likely you are to improve them in the next 12 months. - Has knowledge of algorithms, data structures, and programming paradigms}
\lookupPut{Q19_13}{Rate the following skills by how likely you are to improve them in the next 12 months. - Is available to answer questions}
\lookupPut{Q20}{What factors influence which skills you want to improve on?}
\lookupPut{Q21_1}{Please rate your agreement with each of the following statements: - I know how to improve the OSS-related skills I want to grow.}
\lookupPut{Q21_2}{Please rate your agreement with each of the following statements: - I am taking steps to improve the OSS-related skills I want to grow.}
\lookupPut{Q21_3}{Please rate your agreement with each of the following statements: - I would be more motivated to improve my OSS-related skills if they were displayed publicly on GitHub.}
\lookupPut{Q22_1}{Please rate your agreement with each of the following statements: - I choose to contribute to an OSS project based on the skills that I can gain from the project.}
\lookupPut{Q22_2}{Please rate your agreement with each of the following statements: - I choose to contribute to an OSS project based on the skills that I can offer to the project.}
\lookupPut{Q22_3}{Please rate your agreement with each of the following statements: - I would choose to contribute to a project if I knew it had a collaborator who was strong at a particular skill.}
\lookupPut{Q22_4}{Please rate your agreement with each of the following statements: - I would choose to contribute to a project if I knew they were looking for contributors with a skill I have.}
\lookupPut{Q23}{For the OSS project(s) you maintain, rate the following skills based on how important they are for new contributors to have.}
\lookupPut{Q23_1}{For the OSS project(s) you maintain, rate the following skills based on how important they are for new contributors to have. - Teaches others to be involved in the OSS project}
\lookupPut{Q23_2}{For the OSS project(s) you maintain, rate the following skills based on how important they are for new contributors to have. - Has commitment to the OSS project}
\lookupPut{Q23_3}{For the OSS project(s) you maintain, rate the following skills based on how important they are for new contributors to have. - Has experience with specific tools, frameworks, or programming languages}
\lookupPut{Q23_4}{For the OSS project(s) you maintain, rate the following skills based on how important they are for new contributors to have. - Has strong problem-solving skills}
\lookupPut{Q23_5}{For the OSS project(s) you maintain, rate the following skills based on how important they are for new contributors to have. - Is familiar with OSS practices}
\lookupPut{Q23_6}{For the OSS project(s) you maintain, rate the following skills based on how important they are for new contributors to have. - Puts in effort to make high-quality contributions}
\lookupPut{Q23_7}{For the OSS project(s) you maintain, rate the following skills based on how important they are for new contributors to have. - Is a strong communicator}
\lookupPut{Q23_8}{For the OSS project(s) you maintain, rate the following skills based on how important they are for new contributors to have. - Learns continuously}
\lookupPut{Q23_9}{For the OSS project(s) you maintain, rate the following skills based on how important they are for new contributors to have. - Follows proper processes to make contributions to OSS projects}
\lookupPut{Q23_10}{For the OSS project(s) you maintain, rate the following skills based on how important they are for new contributors to have. - Is a strong project manager}
\lookupPut{Q23_11}{For the OSS project(s) you maintain, rate the following skills based on how important they are for new contributors to have. - Builds good relationships with stakeholders}
\lookupPut{Q23_12}{For the OSS project(s) you maintain, rate the following skills based on how important they are for new contributors to have. - Has knowledge of algorithms, data structures, and programming paradigms}
\lookupPut{Q23_13}{For the OSS project(s) you maintain, rate the following skills based on how important they are for new contributors to have. - Is available to answer questions}
\lookupPut{Q24}{As a maintainer, do you use GitHub to determine what skills a potential contributor has?}
\lookupPut{Q25_1}{As a maintainer, what do you use to determine what skills a potential contributor has? - Selected Choice - Public GitHub activity}
\lookupPut{Q25_2}{As a maintainer, what do you use to determine what skills a potential contributor has? - Selected Choice - Activity on social media platforms}
\lookupPut{Q25_3}{As a maintainer, what do you use to determine what skills a potential contributor has? - Selected Choice - Personal website}
\lookupPut{Q25_4}{As a maintainer, what do you use to determine what skills a potential contributor has? - Selected Choice - Public repositories the contributor owns}
\lookupPut{Q25_5}{As a maintainer, what do you use to determine what skills a potential contributor has? - Selected Choice - Contributions from contributor on your OSS project}
\lookupPut{Q25_6}{As a maintainer, what do you use to determine what skills a potential contributor has? - Selected Choice - Interactions with the potential contributor on your OSS project}
\lookupPut{Q25_7}{As a maintainer, what do you use to determine what skills a potential contributor has? - Selected Choice - Contributions from contributor on other OSS projects}
\lookupPut{Q25_8}{As a maintainer, what do you use to determine what skills a potential contributor has? - Selected Choice - Interactions with the potential contributor on other OSS projects}
\lookupPut{Q25_9}{As a maintainer, what do you use to determine what skills a potential contributor has? - Selected Choice - Other (please specify)}
\lookupPut{Q25_9_TEXT}{As a maintainer, what do you use to determine what skills a potential contributor has? - Other (please specify) - Text}
\lookupPut{Q27}{Consider the feature that shows contributor skills in the outlined red box. Rate your agreement with the following statement: I would like this feature on GitHub.}
\lookupPut{Q28}{Consider the feature that shows contributor skills and their ratings below in the outlined red box. Rate your agreement with the following statement: I would like this feature on GitHub.}
\lookupPut{Q29}{Consider the feature that shows contributor skills and their ratings below in the outlined red box. Rate your agreement with the following statement: I would like this feature on GitHub.}
\lookupPut{Q30}{Consider the feature that shows activities completed that are associated with the skill below in the outlined red box. Rate your agreement with the following statement: I would like this feature on GitHub.}
\lookupPut{Q31}{Consider the feature that shows how to improve contributor skills below. Rate your agreement with the following statement: I would like this feature on GitHub.}
\lookupPut{Q32_1}{Which of the following information about your skills would you display on GitHub to just yourself, your collaborators, and the public? - Skill name}
\lookupPut{Q32_2}{Which of the following information about your skills would you display on GitHub to just yourself, your collaborators, and the public? - Skill rating}
\lookupPut{Q32_3}{Which of the following information about your skills would you display on GitHub to just yourself, your collaborators, and the public? - Which activities you've done that correlate with the skill}
\lookupPut{Q32_4}{Which of the following information about your skills would you display on GitHub to just yourself, your collaborators, and the public? - Ways to improve the skill}
\lookupPut{Q33}{What motivates you to share skills information with others?}
\lookupPut{Q51}{Can you give examples of when it would be helpful to share skills information with others?}
\lookupPut{Q34}{For each of the skills below, rate your agreement with the following statement: if I were proficient at this skill, I would publicly display that I have proficiency in this skill on my GitHub profile.}
\lookupPut{Q34_1}{Skill Ratings For each of the skills below, rate your agreement with the following statement: if I were proficient at this skill, I would publicly display that I have proficiency in this skill on my GitHub profile. - Team player}
\lookupPut{Q34_2}{Skill Ratings For each of the skills below, rate your agreement with the following statement: if I were proficient at this skill, I would publicly display that I have proficiency in this skill on my GitHub profile. - Commitment}
\lookupPut{Q34_3}{Skill Ratings For each of the skills below, rate your agreement with the following statement: if I were proficient at this skill, I would publicly display that I have proficiency in this skill on my GitHub profile. - Familiarity with open-source software}
\lookupPut{Q35}{For each of the programming languages below, rate your agreement with the following statement: if I were proficient at this programming language, I would publicly display that I have proficiency in this programming language on my GitHub profile.}
\lookupPut{Q35_1}{For each of the programming languages below, rate your agreement with the following statement: if I were proficient at this programming language, I would publicly display that I have proficiency in this programming language on my GitHub profile. - JavaScript}
\lookupPut{Q35_2}{For each of the programming languages below, rate your agreement with the following statement: if I were proficient at this programming language, I would publicly display that I have proficiency in this programming language on my GitHub profile. - Python}
\lookupPut{Q35_3}{For each of the programming languages below, rate your agreement with the following statement: if I were proficient at this programming language, I would publicly display that I have proficiency in this programming language on my GitHub profile. - Java}
\lookupPut{Q35_4}{For each of the programming languages below, rate your agreement with the following statement: if I were proficient at this programming language, I would publicly display that I have proficiency in this programming language on my GitHub profile. - PHP}
\lookupPut{Q35_5}{For each of the programming languages below, rate your agreement with the following statement: if I were proficient at this programming language, I would publicly display that I have proficiency in this programming language on my GitHub profile. - C\#}
\lookupPut{Q35_6}{For each of the programming languages below, rate your agreement with the following statement: if I were proficient at this programming language, I would publicly display that I have proficiency in this programming language on my GitHub profile. - TypeScript}
\lookupPut{Q35_7}{For each of the programming languages below, rate your agreement with the following statement: if I were proficient at this programming language, I would publicly display that I have proficiency in this programming language on my GitHub profile. - Shell}
\lookupPut{Q35_8}{For each of the programming languages below, rate your agreement with the following statement: if I were proficient at this programming language, I would publicly display that I have proficiency in this programming language on my GitHub profile. - C}
\lookupPut{Q35_9}{For each of the programming languages below, rate your agreement with the following statement: if I were proficient at this programming language, I would publicly display that I have proficiency in this programming language on my GitHub profile. - Ruby}
\lookupPut{Q36}{Do you have any additional comments about skills in OSS? Is there anything else that you would like to share?}
\lookupPut{MF5}{Please enter your GitHub username.}
\lookupPut{MF6}{Rate your proficiency in each skill below.}
\lookupPut{MF7}{Rate your proficiency in each programming language below.}

%% file: tables/skills-signals-table.tex
\definecolor{cellorange}{rgb}{ 1,  .949,  .8}
\definecolor{cellgreen}{rgb}{ .776,  .878,  .706}
\definecolor{cellblue}{rgb}{ .608,  .761,  .902}
\definecolor{celllightgrey}{rgb}{ .921,  .921,  .921}
\definecolor{celldarkgrey}{rgb}{ .664,  .664,  .664}

\definecolor{shadecolor}{rgb}{.92,  .92, .92}

\begin{table*}[ht]
  \centering

\small

\caption{Skills-signals model based on literature review. [A] denotes an author-defined signal. [CE] denotes a signal from one of three OSS community experts.}
\label{tab:SkillsSignalsTable}
\vspace{-0.5\baselineskip}

    \begin{tabular*}{\linewidth}{p{0.80\linewidth}|p{0.16\linewidth}}

    \hline

             \textbf{Signal} & \hfil \textbf{Literature Source} \\ 

\specialrule{0.1em}{0pt}{0.5pt}
\specialrule{0.1em}{0pt}{0.5pt}

\multicolumn{2}{l}{\cellcolor[rgb]{ .921,  .921,  .921} \textbf{Teaches others to be involved in the OSS project}}\\

\hline

Contributes to a pull request with a newcomer at least 3 times & \hfil \cite{steinmacher2015social} \\
\hline

Comments on code in others' pull requests at least once & \hfil \cite{dias2021makes}\\ 
\hline

Comments overall on others' pull requests at least once & \hfil \cite{dias2021makes}\\
\hline

Contributes at least 5 changes to Markdown files at least once & \hfil \cite{chatterjee2021aid} \cite{steinmacher2015social} \\ 
\hline

Contributes at least 5 changes to community health files at least once & \hfil \cite{github2021setting} \cite{chatterjee2021aid} \cite{steinmacher2015social} \\ 
\hline

Is the only person (aside from the owner) who commented on an issue or pull request at least 3 times & \hfil \cite{dias2021makes} \cite{marlow2013activity} \\
\hline

Has not responded to someone's open issue in less than an hour in the past three months & \hfil \cite{zhou2012make} \\ 
\hline

There is no more than 1 comment on pull requests or issues with toxicity scores over 0.5 in the past year & \hfil \cite{hata2021github} \cite{raman2020stress} \\
\hline

\multicolumn{2}{l}{\cellcolor[rgb]{ .921,  .921,  .921}\textbf{Shows commitment towards the OSS project}}\\
\hline

At least 36 months where there is at least one contribution per month across all projects & \hfil \cite{zhou2012make} [CE] \\ 
\hline

At least 12 months where there is at least one contribution per month across all projects & \hfil \cite{zhou2012make} [CE] \\ 
\hline

Is involved in the discussion of their own PR around 70\% of the time in a particular project in the past year & \hfil \cite{lee2017one} \\
\hline

Number of GitHub members who follow the new contributor is at the 75th percentile across users & \hfil \cite{bao2019large} \cite{eluri2021predicting}\\
\hline

Has write rights to a repository they don't own & \hfil \cite{eluri2021predicting} \\
\hline

Is at the 75th percentile by number of commits to a repository across users & \hfil \cite{eluri2021predicting} \\
\hline

\multicolumn{2}{l}{\cellcolor[rgb]{ .921,  .921,  .921}\textbf{Has knowledge in specific programming languages}}\\
\hline

Has made a commit in the language at least once & \hfil [A]\\
\hline

The lines changed in the language is at the 20th percentile across users & \hfil \cite{bergersen2014construction}  \\
\hline 

The lines changed in the language is at the 40th percentile across users & \hfil \cite{bergersen2014construction}  \\
\hline

The lines changed in the language is at the 60th percentile across users & \hfil \cite{bergersen2014construction}  \\
\hline

The lines changed in the language is at the 80th percentile across users & \hfil \cite{bergersen2014construction}  \\
\hline

\multicolumn{2}{l}{\cellcolor[rgb]{ .921,  .921,  .921}\textbf{Is familiar with OSS practices}}\\
\hline

Has made a commit & \hfil [A]\\ 
\hline

Has opened a pull request & \hfil [A]\\
\hline

Has opened an issue & \hfil [A]\\
\hline

Has made a comment on another's pull requests & \hfil [A]\\ 
\hline

Has made a comment on another's issue & \hfil [A]\\
\hline

Has been assigned to an issue, closed an issue, or has merged a pull request & \hfil [A]\\
\hline

\end{tabular*}
\end{table*}

%% file: tables/skills-importance.tex
\begin{table}
\centering
\small

\caption{Participants who found each skill important.}
\label{tab:SkillsImportanceTable}
\vspace{-0,5\baselineskip}
\begin{tabular*}{\linewidth}{p{0.70\linewidth}|p{0.20\linewidth}}
\hline

\textbf{Skill} &  \hfil  \textbf{Importance} \\

\specialrule{0.1em}{0pt}{0.5pt}
\specialrule{0.1em}{0pt}{0.5pt}
Teaches others to be involved in the OSS project & \hfil 64\% \\
\hline

Shows commitment towards the OSS project & \hfil 67\% \\
\hline

Has knowledge in specific programming languages & \hfil 45\% \\
\hline

Is familiar with OSS practices & \hfil 56\% \\
\hline

\end{tabular*}
\end{table}

%% file: tables/merged.tex
\begin{table}
  \centering

\small

\newlength{\myDistLength}
\settowidth{\myDistLength}{Distribution}  

\def\mybarchart#1#2#3#4#5#6{
\resizebox {\the\myDistLength} {7.5pt} {%
\begin{tikzpicture}[]
\begin{axis}[
      axis background/.style={fill=gray!30, draw=gray!30},
      axis line style={draw=none},
      tick style={draw=none},
      ytick=\empty,
      xtick=\empty,
      ymin=0, ymax=0.70, %
      xmin=0, xmax=6]
\addplot [
      ybar interval=.5,
      fill=blue,
      draw=none,
]
	coordinates {(6,#6) (5,#5) (4,#4) (3,#3) (2,#2) (1,#1)}; %
\addplot [
      ybar interval=.5,
      fill=orange,
      draw=none,
]
	coordinates {(1,#1) (0,0.30)}; %
\end{axis}%
\end{tikzpicture}%
}%
}

\caption{Results of the survey and evaluation of \toolname. Mean values in the survey are from people who reported to have the skill. The tool's precision scores are from those who had self-evaluated
their skill level to be greater than 0 and 3.}
\label{tab:MergedTable}
\vspace{-0.5\baselineskip}
\begin{tabular}{@{}l@{\phantom{X}}c@{\phantom{X}}c@{\phantom{X}}cc@{\phantom{X}}c@{\phantom{X}}c@{}}
\toprule
& \multicolumn{3}{c}{Survey} & \multicolumn{3}{c}{Tool} \\
\cmidrule(lr){2-4}\cmidrule(lr){5-7}
Skill & \stackanchor{Distribution}{\makebox[\the\myDistLength][s]{\color{orange}0 \color{blue}1 2 3 4 5}} & \color{blue} Mean & \stackanchor{Would}{display} & \stackanchor{Distribution}{\makebox[\the\myDistLength][s]{\color{orange}0 \color{blue}1 2 3 4 5}} & \stackanchor{Prec.}{(>0)} & \stackanchor{Prec.}{(>3)} \\
\midrule
OSSFamiliarity & \mybarchart{0.0253164556962025}{0.00949367088607595}{0.0348101265822785}{0.208860759493671}{0.357594936708861}{0.363924050632911} & 4.06 & 79\% & \mybarchart{0}{0.00316455696202532}{0}{0.0158227848101266}{0.0474683544303797}{0.933544303797468} & \textbf{97\%} & 74\% \\
Commitment & \mybarchart{0.0411392405063291}{0.0221518987341772}{0.0664556962025316}{0.174050632911392}{0.395569620253165}{0.300632911392405} & 3.92 & 67\% & \mybarchart{0.0632911392405063}{0.484177215189873}{0.243670886075949}{0.104430379746835}{0.069620253164557}{0.0348101265822785} & 96\% & 63\% \\
TeamPlayer & \mybarchart{0.0284810126582278}{0.00949367088607595}{0.0632911392405063}{0.180379746835443}{0.436708860759494}{0.281645569620253} & 3.94 & 69\% & \mybarchart{0}{0}{0.189873417721519}{0.272151898734177}{0.484177215189873}{0.0537974683544304} & \textbf{97\%} & 73\% \\
\midrule 
Javascript & \mybarchart{0.0664556962025316}{0.126582278481013}{0.104430379746835}{0.221518987341772}{0.246835443037975}{0.234177215189873} & 3.38 & \textbf{80\%} & \mybarchart{0.319620253164557}{0.139240506329114}{0.246835443037975}{0.117088607594937}{0.069620253164557}{0.107594936708861} & 95\% & 65\% \\
Python & \mybarchart{0.107594936708861}{0.107594936708861}{0.0917721518987342}{0.215189873417722}{0.243670886075949}{0.234177215189873} & 3.45 & \textbf{80\%} & \mybarchart{0.417721518987342}{0.120253164556962}{0.060126582278481}{0.240506329113924}{0.0443037974683544}{0.117088607594937} & 96\% & 88\% \\
Java & \mybarchart{0.215189873417722}{0.199367088607595}{0.158227848101266}{0.161392405063291}{0.129746835443038}{0.136075949367089} & 2.80 & 67\% & \mybarchart{0.70253164556962}{0.0443037974683544}{0.0759493670886076}{0.069620253164557}{0.0316455696202532}{0.0759493670886076} & 96\% & 67\% \\
Php & \mybarchart{0.40506329113924}{0.189873417721519}{0.123417721518987}{0.129746835443038}{0.0854430379746835}{0.0664556962025316} & 2.52 & 54\% & \mybarchart{0.806962025316456}{0.110759493670886}{0.0158227848101266}{0.00949367088607595}{0.0126582278481013}{0.0443037974683544} & 89\% & 56\% \\
C\# & \mybarchart{0.462025316455696}{0.180379746835443}{0.107594936708861}{0.107594936708861}{0.0632911392405063}{0.0791139240506329} & 2.54 & 64\% & \mybarchart{0.867088607594937}{0.120253164556962}{0.00632911392405063}{0}{0.00316455696202532}{0.00316455696202532} & 83\% & \textbf{100\%} \\
Typescript & \mybarchart{0.322784810126582}{0.145569620253165}{0.107594936708861}{0.145569620253165}{0.174050632911392}{0.104430379746835} & 2.98 & 73\% & \mybarchart{0.721518987341772}{0.113924050632911}{0.0506329113924051}{0.0379746835443038}{0.060126582278481}{0.0158227848101266} & 92\% & 79\% \\
Shell & \mybarchart{0.113924050632911}{0.129746835443038}{0.123417721518987}{0.278481012658228}{0.215189873417722}{0.139240506329114} & 3.13 & 69\% & \mybarchart{0.518987341772152}{0.113924050632911}{0.0411392405063291}{0.0854430379746835}{0.113924050632911}{0.126582278481013} & 91\% & 61\% \\
C & \mybarchart{0.243670886075949}{0.186708860759494}{0.126582278481013}{0.170886075949367}{0.129746835443038}{0.142405063291139} & 2.89 & 73\% & \mybarchart{0.832278481012658}{0.0759493670886076}{0.0284810126582278}{0.0253164556962025}{0.0158227848101266}{0.0221518987341772} & 98\% & 92\% \\
Ruby & \mybarchart{0.582278481012658}{0.15506329113924}{0.0791139240506329}{0.0791139240506329}{0.0443037974683544}{0.060126582278481} & 2.46 & 66\% & \mybarchart{0.794303797468354}{0.10126582278481}{0.00949367088607595}{0.0569620253164557}{0.0379746835443038}{0} & 77\% & 58\% \\
\bottomrule
\end{tabular}

\end{table}